\documentclass[onecolumn,authoryear]{els-mrw}

\usepackage{amsmath,amssymb,amsfonts,amsthm,makeidx,graphicx}
\usepackage{txfonts}
\usepackage{helvet}
\usepackage{subfig, siunitx}

\usepackage{amsmath,amssymb}

\begin{document}

\def\spitzer{\textit{Spitzer}}
\def\herschel{\textit{Herschel}}
\def\xmm{\textit{XMM-Newton}}
\def\wmap{\textit{WMAP}}
\def\planck{\textit{Planck}}
\def\litebird{\textit{LiteBird}}
\def\prism{\textit{PRISM}}
\def\pixie{\textit{PIXIE}}
\def\cmbpol{\textit{CMBpol}}
\def\ebex{\textit{EBEX}}
\def\apexsz{\textit{APEX-SZ}}
\def\wise{\textit{WISE}}
\def\iras{\textit{IRAS}}
\def\aap{A\&A}
\def\apss{Ap\&SS}
\def\araa{ARA\&A}
\def\apj{ApJ}
\def\mnras{MNRAS}
\def\nat{Nature}
\def\na{New Astronomy}
\def\apjl{ApJL}
\def\apjs{ApJS}
\def\prd{PRD}
\def\jcap{JCAP}
\def\prl{PRL}
\def\plb{PLB}
\def\physrep{Physics Reports}
\newcommand{\desi}{DESI}
\newcommand{\vista}{{\sc VISTA}}
\newcommand{\spt}{SPT}
\newcommand{\sptsz}{SPT-SZ}
\newcommand{\sptpol}{SPTpol}
\newcommand{\sptnew}{SPT-3G}
\newcommand{\pb}{POLARBEAR}
\newcommand{\pbb}{POLARBEAR2}
\newcommand{\pbnew}{Simons Array}
\newcommand{\so}{Simons Observatory}
\def\great#1{\textbf{#1}}
\def\good#1{\textbf{\textcolor{orange}{#1}}}
\def\ok#1{\textcolor{black}{#1}}
\def\update#1{\textbf{\textcolor{red}{#1}}}
\newcommand{\biceptwo}{BICEP2}
\newcommand{\bicepthree}{BICEP3}
\newcommand{\spud}{SPUD}
\newcommand{\keckarray}{KECK Array}
\newcommand{\abs}{ABS}
\newcommand{\act}{ACT}
\newcommand{\actpol}{ACTpol}
\newcommand{\actnew}{AdvancedACT}
\newcommand{\spider}{SPIDER}
\newcommand{\msolar}{\ensuremath{M_\odot}}
\newcommand{\lcdm}{\ensuremath{\Lambda{\rm CDM}}}
\newcommand{\wcdm}{\ensuremath{{\rm wCDM}}}
\newcommand{\sigeight}{\ensuremath{\sigma_8}}
\newcommand{\ho}{\ensuremath{H_0}}
\newcommand{\bsz}{\ensuremath{B_{SZ}}}
\newcommand{\ttsz}{\ensuremath{T_{SZ}}}
\newcommand{\Yx}{\ensuremath{Y_{X}}}
\newcommand{\yhe}{\ensuremath{Y_{\rm He}}}
\newcommand{\summnu}{\ensuremath{\sum m_\nu}}
\newcommand{\sqdeg}{\ensuremath{{\rm deg^2}}}
\newcommand{\ukarcmin}{\ensuremath{\mu{\rm K-arcmin}}}
\newcommand{\uks}{\ensuremath{\mu{\rm K \sqrt{s}}}}
\def\summnu{\ensuremath{\Sigma m_\nu}}
\def\neff{\ensuremath{N_\mathrm{eff}}}
\def\nrun{\ensuremath{n_\mathrm{run}}}
\def\emode{$E$-mode}
\def\emodes{$E$-modes}
\def\bmode{$B$-mode}
\def\bmodes{$B$-modes}
\def\omm{\ensuremath{\Omega_m}}
\def\ombh{\ensuremath{\Omega_b h^2}}
\def\omch{\ensuremath{\Omega_c h^2}}
\def\omdmh{\ensuremath{\Omega_{dm} h^2}}
\newcommand{\erosita}{eROSITA}
\newcommand{\boss}{BOSS}
\newcommand\add[1]{\textbf{TBD: #1}}
\newcommand\tbd[1]{\textcolor{red}{#1}}
\newcommand\comment[1]{}
\makeatletter
\newcommand\thefontsize[1]{{#1 The current font size is: \f@size pt\par}}
\makeatother

\newcommand{\des}{{DES}}
\newcommand{\more}{\textbf{MORE}}
\chapter{Polarization of the Cosmic Microwave Background}\label{chap1}

\author[1]{Mahsa Rahimi}%
\author[1]{Christian L. Reichardt}%

\address[1]{\orgname{The University of Melbourne}, \orgdiv{School of Physics}, \orgaddress{Parkville VIC 3010, Australia}}

\articletag{v 0.1}

\maketitle

\begin{glossary}[Nomenclature]
\begin{tabular}{@{}lp{34pc}@{}}
CMB &Cosmic microwave background\\
TES &Transition edge sensor\\
\end{tabular}
\end{glossary}

\begin{abstract}[Abstract]

Since the first detection by the DASI experiment in 2002, measurements of the polarization of the cosmic microwave background (CMB) have grown into an important role in
testing our understanding of conditions in the early universe and cosmology.
The field has seen rapid experimental progress, driven in large part by the desire to make increasingly precise measurements of CMB polarization.
Precise measurements of the CMB polarization anisotropies contain as much information as the CMB temperature anisotropy, and promise to unlock new tests of physics and the standard cosmological model.

In this chapter, we first discuss how polarization is produced through Thomson scattering, after which types of polarization patterns are connected to the cosmological sources.
Finally, we briefly discuss the experimental hardware that enables these measurements.
\end{abstract}

\begin{BoxTypeA}[keypoints]{Key points}
\begin{itemize}
\item \textbf{Cosmic Microwave Background (CMB)} is 2.7\,K black body radiation from the early Universe. The CMB radiation is partially polarized by Thomson scattering with electrons, giving rise to CMB polarization.
\item \textbf{CMB anisotropies} are tiny spatial variations in the temperature and polarization patterns of the CMB radiation.
\item \textbf{Thomson scattering} is elastic scattering between photons and a charged particle (generally electrons).
\item \textbf{Radiation quadrupoles} are a radiation pattern with more intense incoming radiation on one axis, and less intense radiation on the perpendicular axis.
\item \textbf{Inflation} is a period of accelerating expansion in the first moments of the Universe, during which the Universe is theorized to have expanded by a factor of $e^{60} - e^{100}$. Inflation is expected to produce density perturbations, which have been observed, and gravitational waves, which have yet to be observed.
\item \textbf{Last scattering surface} is an image of the last time most CMB photons have scattered off matter. Last scattering occurs at recombination in the early Universe when the temperature cools enough for electrons and protons to recombine into neutral Hydrogen.
\item \textbf{Linearly polarized} light has a preferred direction for the electric fields of the electromagnetic wave.
\item \textbf{E-mode polarization} is a pattern of linear polarization which has  zero curl. E-mode polarization is produced by acoustic waves in the CMB last scattering surface.
\item \textbf{B-mode polarization} is a pattern of linear polarization which has  zero gradient. B-mode polarization is produced by gravitational waves in the CMB last scattering surface.
\item \textbf{Bolometers} are devices to measure radiative power through a temperature-dependent resistance.
\item \textbf{Transition edge sensors (TESes)} are sensitive thermometers that utilize the sharp change in resistance as a superconductor transitions from normal to superconducting states.
\end{itemize}

\end{BoxTypeA}

\section{Introduction}\label{chap1:sec1}

The cosmic microwave background (CMB) radiation is relic radiation from the hot plasma of the early Universe, some 400,000 years after the Big Bang.
The CMB photons decoupled from electrons at recombination when the Universe had cooled sufficiently for the electron-proton plasma to combine to neutral hydrogen.
This occurred at a temperature of 3000\,K, with the subsequent expansion of the Universe by a factor of 1100 lowering the CMB temperature today to $2.72548 \pm 0.00057$ \citep{fixsen09}.
As the scattering cross-section of hydrogen is small compared to the Thomson scattering cross-section of a free electron, the mean-free path of a photon increases precipitously during recombination.
Most CMB photons have not scattered in the history of the Universe since recombination, which is also called the last scattering surface.

While scattering in a perfectly homogenous distribution of matter and photons does not create polarization, the early Universe is very slightly inhomogenous.
In inflationary models, the initial perturbations are produced by quantum fluctuations in the inflationary field.
These initial seeds then grow under the influence of gravity to the galaxies and large scale structure we see today.
CMB anisotropy measurements provide a snapshot of the density perturbations at the last scattering surface.

Density (scalar) perturbations in the early hot plasma produce both temperature and polarization anisotropies in the CMB.
As detailed in \S\ref{sec:gen}, the scattered light from a quadrupole radiation field scattering off a free electron will have a net linear polarization.
The growth of quadrupole moments is suppressed while the photon mean free path is small, thus the resulting polarization patterns are more sensitive to physics occurring as the mean free path grows in the lead up to last scattering.
Measuring polarization as well as temperature anisotropy roughly doubles the information content that can be extracted from CMB power spectrum measurements \citep{scott16}, and can break parameter degeneracies in some cosmological models \citep{galli14}.
Precision measurements of CMB polarization can better constrain physics of the early Universe, for instance by searching for possible new particles produced during the hot Big Bang \citep{abazajian16}.
Additionally, by more tightly constraining starting conditions in the early Universe, CMB polarization measurements strengthen observational tests of the Universe's late-time evolution.

The same situation also occurs in the late universe when light from the first stars and galaxies reionizes the Universe.
The large-scale CMB quadrupole scatters off free electrons in the late Universe, adding additional polarization power at scales comparable to the horizon size at reionization.
This ``reionization bump" shows up at very large angular scales ($\ell\lesssim 10$) and measurements of the reionization bump by the WMAP and Planck satellites have been used to estimate the optical depth of the Universe to reionization \citep[e.g.,][]{wmap9params, planck18_6}.

Finally, the standard model of cosmology includes a brief period of very rapid expansion  in which the size of the Universe expands by a factor of $e^{60} - e^{100}$, called inflation.
Models of inflation have made a number of successful predictions \citep{guth81,guth82}, such as that the Universe would be flat with a nearly scale-invariant power spectrum of initial density perturbations.
The exponential expansion of the Universe during inflation is expected to produce a background of gravitational waves (tensor perturbations), which has not yet been observed.
The power in these gravitational waves is related to the energy scale of inflation, and a detection (or upper limit) on these waves is one way to constrain the physics of inflation.
The stretching and squeezing of spacetime at the last scattering surface by gravitational waves will also produce quadrupole radiation fields that scatter off the free electrons to produce linear polarization patterns.
As discussed in \S\ref{sec:sources}, the symmetry of the polarization patterns produced by gravitational waves and density perturbations is different, so the physical origins of the observed polarization are distinguishable in principle.
 The search for these primordial gravitational waves from inflation has been described as a search for the ``smoking gun of inflation".

\section{Generating CMB Polarization }\label{sec:gen}

\begin{figure}
    \centering
    \includegraphics[scale=0.4]{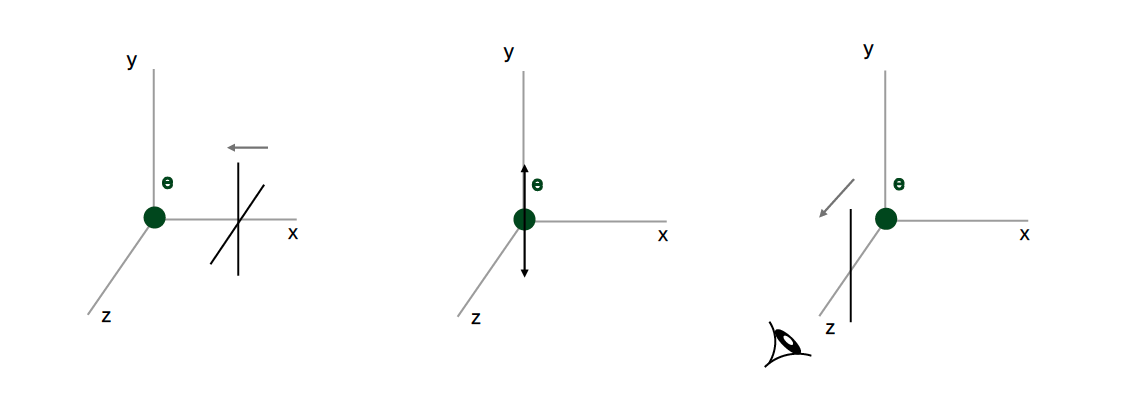}
    \caption{The scattered light is polarized for Thomson scattering of light off a free electron.
  Left: Unpolarized light (no preferred direction for the electric field) is incident upon the electron from the +x direction.
  Middle: The electromagnetic fields oscillate the electron in the plane perpendicular to propagation, here in the yz-plane. Viewed from the +z-axis, the electron is only moving in $\pm$ y.
  Right: An observer on the z-axis sees scattered light that is linearly polarized along the y-axis (illustration inspired from \citet{Hu97}).}
    \label{fig:scattering}
\end{figure}

\begin{figure}
    \centering
    \includegraphics[scale=0.4]{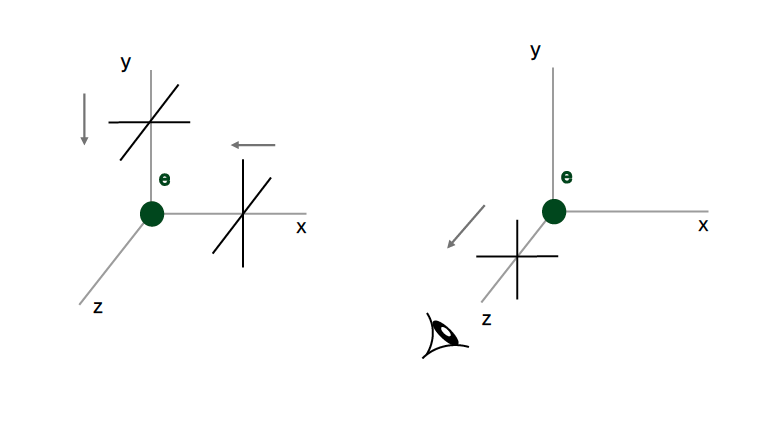}
    \caption{For an isotropic distribution of light, the scattered light from Thomson scattering is unpolarized.
 Left: For an isotropic distribution of light, the incoming intensity from the +x and +y directions (or any other two directions separated by 90$^\circ$) will be equal.
 Viewed by an observer along  the z-axis, the electron will oscillate equally in the x and y directions.
 Right: Thus there will be no preferred direction for the electric field of the scattered light; the light is unpolarized. (illustration inspired from \citet{Hu97}).}
    \label{fig:isotropic}
\end{figure}

The scattering mechanism between CMB photons and a plasma is Thomson scattering off the free electrons.
Thomson scattering will produce linearly polarized light when the incident radiation field has a quadrupole moment.
To understand the origin of polarization and when polarization is created, first consider a EM plane wave incident upon an electron from the $+x$ direction (see Fig.~\ref{fig:scattering}).
The electron will oscillate in the $y-z$ plane.
An observer in the $+z$ direction will see linearly polarized scattered light (in the $y$ direction) since the electron does not move in $x$.
This is over-simplified of course, as one would not expect all light to come from the same direction.
In the case of an isotropic radiation field (Fig.~\ref{fig:isotropic}), one has equal intensities of light impinging on the electron from all directions in the $x-y$ plane, and the electron oscillates in all directions equally.
Thus the net linear polarization of scattered light would be zero since there is no preferred direction.

\begin{figure}
    \centering
    \includegraphics[scale=0.4]{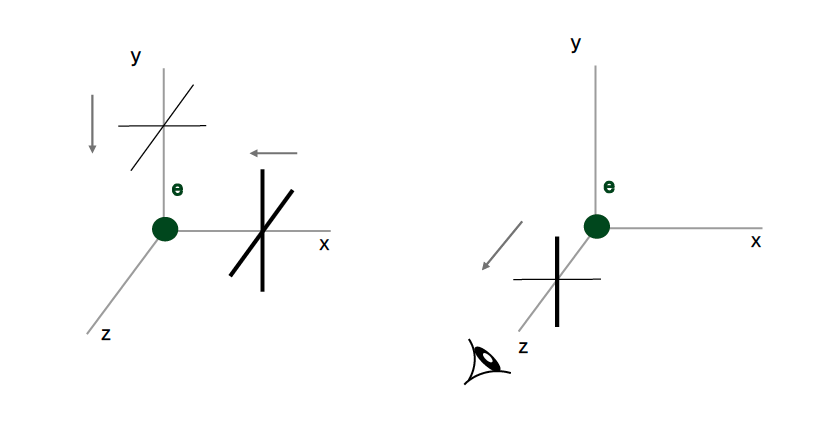}
    \caption{
  A quadrupole radiation field generates linear polarization in the light scattered off a free electron.
 Left:  A quadrupole radiation field means the light has a higher intensity (hotter) along one axis (taken to be the x-axis here), and lower intensity (cooler) on the perpendicular axis (here the y-axis).
 Right: Viewed by an observer along the z-axis, the electron's oscillations will be larger in the y-direction than x-direction.
 The electric field of the scattered light will tend to be larger in the y-direction, and the scattered light will be partially polarized (illustration inspired from \citet{Hu97}).}
    \label{fig:quad}
\end{figure}

Now consider the case where there is a radiation quadrupole (Fig.~\ref{fig:quad}).
Without loss of generality, we can take the incoming radiation to have a higher intensity (higher temperature) in the $\pm y$ directions than in the $\pm x$ directions.
The induced oscillations in the electron then will be larger on average along the $x$-direction than the $y$-direction.
This causes the scattered light off the electron towards our observer in the $+z$ direction to be partially linearly polarized in the $x$-direction.
An alternative way to understand this is that the differential cross-section of Thomson scattering, $\frac{d\sigma}{d\Omega}$, defined as the radiated intensity per solid angle $d\Omega$ over the incoming intensity per solid angle $d\Omega$, is proportional to the scalar product of the incident and scattered polarization vectors:
\begin{equation}
\frac{d\sigma}{d\Omega} = \frac{3\sigma_T}{8\pi} |\hat{\varepsilon}^\prime \cdot \hat{\varepsilon}|^2,
\end{equation}
where $\sigma_T$ is the Thomson cross-section and the unit vectors $\hat{\varepsilon}^\prime,  \hat{\varepsilon}$ are aligned with the polarization direction in the incoming and outgoing light respectively.
This means that the intensity of the scattered light peaks when the polarization of the incident and scattered light are parallel (For a pedagogical review, refer to \cite{2020moco.book.....D}, \cite{Hu97}).
Importantly, Thomson scattering can not produce circular polarization.
Thus the CMB is not expected to be circularly polarized, and most CMB experiments are insensitive to circular polarization.
A larger quadrupole moment will increase the degree of linear polarization, with the scattered light 100\% polarized when there is no incident light from the `cold' directions.
The CMB anisotropies are approximately 10\% polarized due to this mechanism.

Well before recombination, photons are tightly coupled to the electrons of the plasma, and the photon mean free path between scattering is very short.
With each scattering event, the photon direction is randomized which suppresses the growth of radiation quadrupoles.
It is only near the last scattering surface that the photon mean free path lengthens sufficiently to allow quadrupole anisotropies to begin to grow.
This is in contrast to the dark matter density perturbations which grow under the influence of gravity once a mode enters the horizon.
Thus relative to the CMB temperature anisotropy, CMB polarization anisotropies are more sensitive to the details of what is happening around the time of recombination.

\section{Sources of Polarization } \label{sec:sources}

As we have seen, quadrupole anisotropy in the radiation can yield polarized scattered light.
In a spherical harmonics basis, a quadrupole corresponds to $\ell=2$, which may have multipole values of $m=0,1,2$.
The different quadrupoles are produced by scalar (spin-0), vector and tensor (spin-2) perturbations respectively.
Each type of quadrupole (and generating perturbation) yields a different polarization pattern and potential correlation between temperature and polarization.

Observationally, we have clear evidence for primordial density (scalar) perturbations in the Universe, as the density perturbations are the seeds of all the structures we see in the Universe.
We have not detected any direct evidence as yet for primordial vector or tensor (gravitational waves) perturbations.
As noted earlier, inflationary models predict some level of primordial tensor perturbations (the inflationary gravitational waves).


\subparagraph{Density (Scalar) Perturbations}

\begin{figure}[htbp]
\centering
\subfloat[scalar quadrupole moment]
{\includegraphics[width=0.3\textwidth]{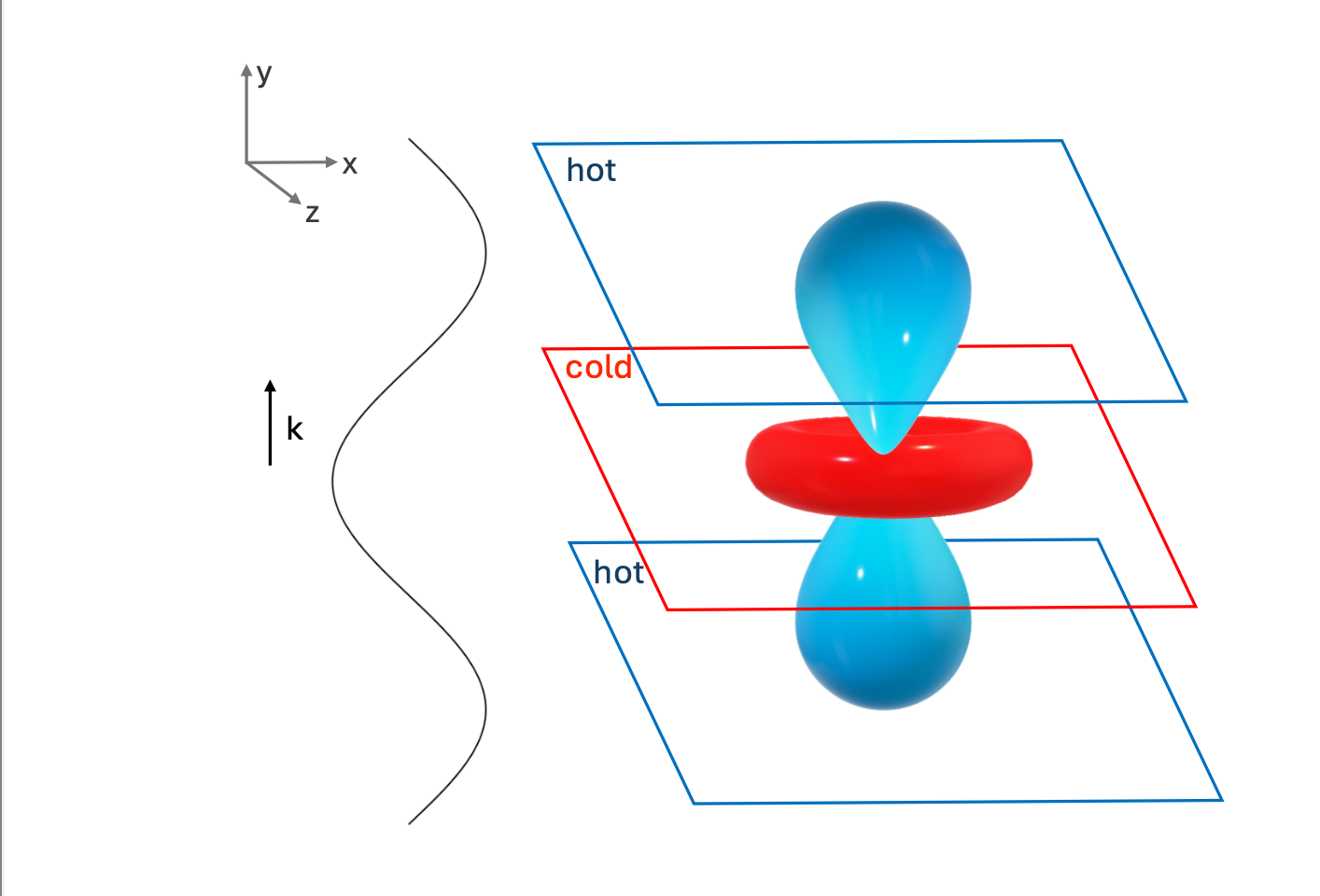}
\label{fig:quadrupole_m0}}
\subfloat[vector quadrupole moment]
{\includegraphics[width=0.3\textwidth]{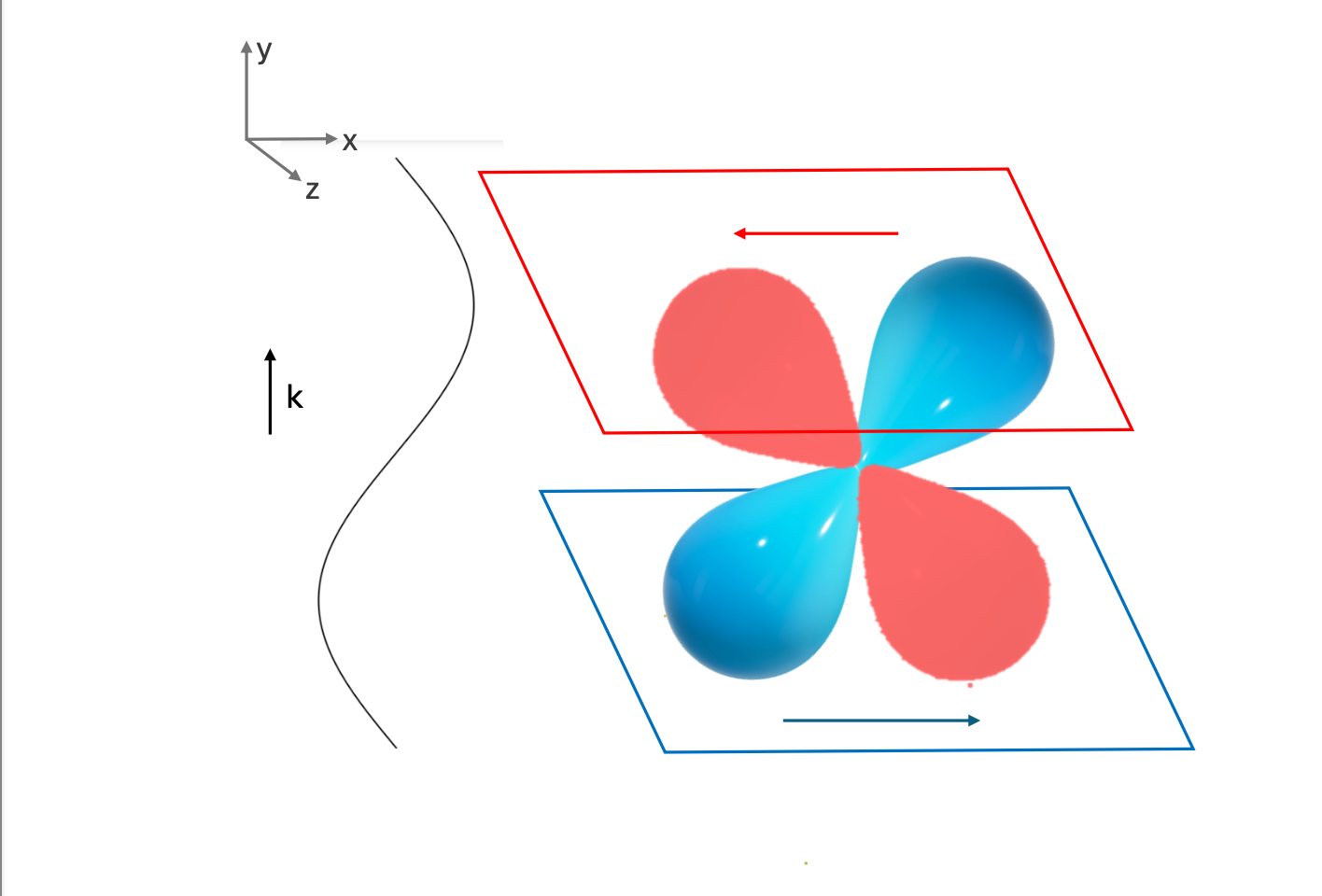}
\label{fig:quadrupole_m1}}
\subfloat[tensor quadrupole moment]
{\includegraphics[width=0.3\textwidth]{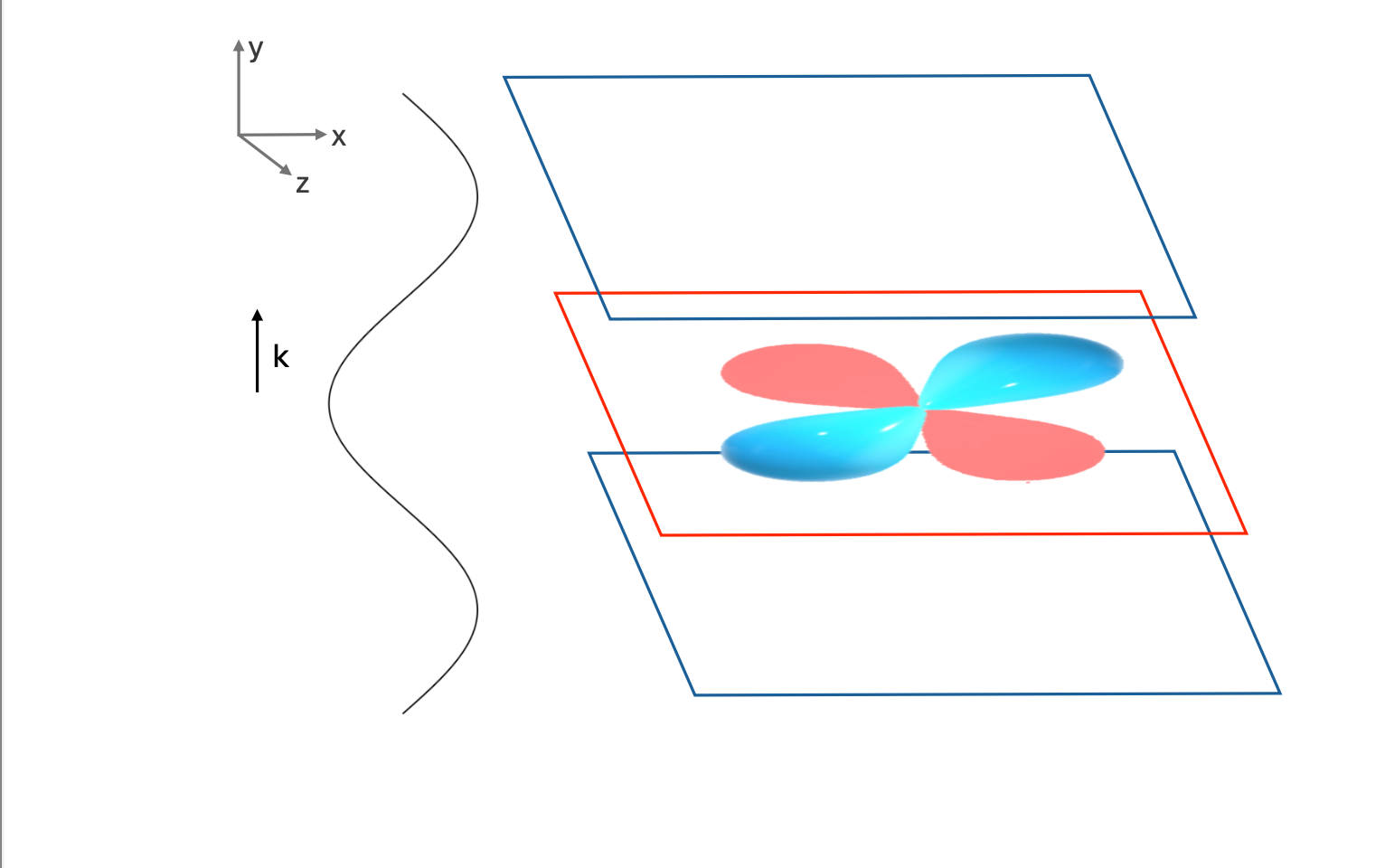}
\label{fig:quadrupole_m2}}
\caption{The quadrupole moments of scalar (a), vector (b) and tensor (c) perturbations, representing $m=0$, $m=1$, $m=2$ states of quadrupole, respectively. The blue/red planes represent the hot/cold or crest/trough planes of a single plane wave. The blue (red) lobes show the hot (cold) regions. The vectors point to the fluid velocity direction. Illustration adapted from \cite{Hu97}.}
\label{fig:quadrupoles}
\end{figure}

One of the successes of CMB observations has been to measure the spectrum of primordial perturbations and show these initial conditions are consistent with a nearly-scale invariant power spectrum of density (scalar) perturbations.
The tiny initial perturbations in density grow under the influence of gravity to produce first the CMB anisotropy and eventually all the structures we observe in the universe.
Density perturbations can create the $m=0$ version of a radiation quadrupole.
Imagine an acoustic plane wave traveling in the y-direction perpendicular to the line of sight ($\hat{z}$) through the plasma, as in Fig.~ \ref{fig:quadrupole_m0}.
The passing wave will compress and rarify the plasma, creating a series of hot (denser) and cold (less dense) planes.
An electron at a hot peak will see more radiation in the $\pm x$ directions and less radiation in the $\pm y$ directions, leading to  linear polarization in the $y$-direction (parallel to the plane wave's direction of propagation).
These directions are flipped for electrons at a cold peak, leading to linear polarization perpendicular to the plane wave's propagation direction.
The local radiation field around an electron at the null of the wave is a gradient rather than a quadrupole, so there is no polarization at the nulls.
Thus acoustic waves in the plasma will create polarization either parallel or perpendicular to the wave direction.
It is also important to note that the largest polarization signal occurs at the peaks of the temperature signal, so scalar perturbations produce a non-zero correlation between the temperature and polarization anisotropies.

\subparagraph{Vorticity or Vector Perturbations}

Vector perturbations represent vortical flows of matter, where the velocity field has a zero gradient and non-zero curl.
A key difference to the last case is that while gravity amplifies scalar perturbations, any vector perturbations will decay in an expanding universe as the inverse square of the scale factor.
Simple models of inflation do not produce vector perturbations, and even in more complicated inflationary scenarios the vector perturbations will have decayed away before recombination.
However, observable vector perturbations may be generated in some models, for instance by topological defects near recombination.
Note that while the vector perturbations will continue to decay away after recombination, their imprint (if any) on the CMB anisotropies would be preserved.
Observational evidence for vector perturbations has not yet been uncovered.

We consider a single Fourier mode of the Fourier representation of the vector perturbations. 
In real space, this will generate velocity fields towards $\pm \hat{x}$ with magnitudes that vary along $\hat{y}$ with the phase of the Fourier mode, as in Fig.~\ref{fig:quadrupole_m1}.
We note that while this picture has been chosen to resemble the acoustic and gravitational waves in the left and right panels of the figure, vector perturbations do not satisfy a wave equation and do not oscillate unlike the other two cases.
The direction of velocity reverses between each peak.
An electron at the null will see a blueshift (hotter) then redshift (cooler) for the flow to the left above it, and a redshift (cooler) then blueshift (hotter) for the flow to the right below it.
This quadrupole pattern corresponds to the $\ell=2, m=1$ multipole moment.
The polarization produced will be largest at the nulls and at an angle $\pm45^\circ$ to the wave direction.




\subparagraph{Gravitational waves or Tensor Perturbations}

Tensor perturbations, or gravitational waves, are distortions in the spacetime metric.
A purely exponential expansion during inflation will lead to a scale-invariant initial power spectrum of these gravitational waves on super-horizon scales, and inflation models generically predict inflationary gravitational waves will exist with a nearly scale-invariant spectrum \citep{durrer05}.
For historical reasons, the amplitude of the gravitational wave spectrum is parameterized as the ratio of the tensor to scalar power at some angular scale, normally called $r$.
A major thrust of current research is to search for these inflationary gravitational waves.

 As a tranverse wave, tensor perturbations compress and stretch the spacetime in directions perpendicular to the wave (see Fig.~\ref{fig:quadrupole_m2}).
 There are two possible polarizations of a gravitational wave, which we will label the $+$ and $\times$ polarizations.
 Consider a circular group of test particles with a gravitational wave propagating in the $\hat{y}$ direction normal to the $xz$-plane of the circle.
  The $+$ polarization alternately compresses and stretches the spacetime in horizontal and vertical directions causing the circle to be alternately elongated in top-to-bottom and side-to-side directions.
  The $\times$ polarization distorts the spacetime in a $45^{\circ}$ alignment with respect to the $+$ polarization.
 The rarefication and compression of the spacetime translates to redshift and blueshift of the photons in these directions, representing a temperature quadrupole moment with $\ell=2,m=2$, as shown in Figure \ref{fig:quadrupole_m2}.
The largest polarization signal will be seen by an observer whose line of sight is parallel to the wave direction.



\section{Representations of CMB polarization}

\subsection{Stokes Parameters}

The Stokes parameters, $\{I, Q, U, V=0\}$,  describe the intensity and polarization of light.
The intensity $I$ is the light's total intensity, while $Q$ and $U$ are measures of the light's linear polarization.
$V$ is a measure of the circular polarization of the radiation.
As noted in \S\ref{sec:gen}, Thomson scattering does not produce circular polarization, so there should be no cosmological circular polarization.
We expect $V=0$ for the CMB, and will drop $V$ for the remainder of the chapter.

The linear polarization parameters $Q$ and $U$ are the key parameters for CMB polarization.
$Q>0 ~(Q<0)$ represents polarization in the N-S (E-W) directions, while $V>0~(V<0)$ refers to NE-SW~(NW-SE) directions.
Mathematically, considering the components of the electric field of the radiation along either the x/y axes ($E_x, E_y$) or the a/b axes ($E_b, E_b$)  which are rotated 45$^\circ$ from the x/y axes, the Stokes parameters are
\begin{equation}
\begin{aligned}
     I = |E_x|^2 + |E_y|^2 \\
     = |E_a|^2 + |E_b|^2, \\
    Q = |E_x|^2 - |E_y|^2, \\
    U = |E_a|^2 - |E_b|^2.
\end{aligned}
\end{equation}


One can define a complex vector, P, representing the total linear polarization:
\begin{equation}
    P = Q + iU
\end{equation}
which transforms like a spin-2 field under rotation (i.e. it is a double-sided arrow and by $180^{\circ}$ rotation remains the same).
The magnitude and angle of polarization are then defined as:
\begin{equation}
\begin{aligned}
    |P| = \sqrt{Q^2+U^2} , \\
    \theta = \frac{1}{2} \arctan(\frac{Q}{U})
\end{aligned}
\end{equation}
with $\theta$ measured clockwise from the North direction.

It is important to note that these definitions are not invariant to rotations and changes to the local coordinate system.
Specifically, if one rotates the axes by an angle $\psi$, the $Q$ and $U$ parameters will transform as:
\begin{equation}
\begin{aligned}
Q^\prime = Q cos 2\psi + U sin 2 \psi, \\
U^\prime = - Q sin 2\psi + U cos 2 \psi,
\end{aligned}
\end{equation}
while the I parameter is unchanged.
In CMB experiments, Stokes parameters are generally used when creating temperature and polarization maps from detectors where there are natural directions defined by the experiment's geometry.

\subsection{$E$-modes and $B$-modes}

 For cosmological analyses, $I$, $Q$ and $U$ maps are generally converted to $T$, $E$ and $B$ maps, as these are (1)  invariant under translations and rotations and (2) can be connected to the underlying physical sources of anisotropy.
 The polarization terms, $E$ and $B$, are respectively a scalar ($E$) and pseudo-scalar ($B$) field  (for more details, refer to \cite{RN1123}).
The nomenclature of $E$- and $B$-modes refers to the gradient-like and curl-like nature of the electric (E) and magnetic (B) fields.
$E$-modes have a gradient and zero curl, while $B$-modes have zero gradient and non-zero curl.
Alternatively, one can see the $E$- and $B$-modes are the parity eigenstates of polarization.
$E$-modes have even parity (do not change sign under a parity transformation) while $B$-modes have odd parity (change sign under a parity transformation).

A full sky Q and U polarization map can be decomposed in spherical harmonic space using the spin-weighted spherical harmonics $ _{\pm2}Y_{\ell m}$  as:
\begin{equation}
    (Q \pm iU) (\hat{n}) = \sum_{\ell}\sum_{m} (E_{\ell m} \pm i B_{\ell m })  _{\pm2}Y_{\ell m}(\hat{n}) \\
\end{equation}
The coefficients of this decomposition are the $E$- and $B$-modes.
While the transformation from $Q$/$U$ to $E$/$B$ is invertible and linear on the full sky, this transformation is non-local.
Actual CMB experiments observe only part of the sky due to the Milky Way and/or the survey area, which can introduce confusion between $E$- and $B$- modes (read \citet{smith06} for more discussion of this point and approaches to handle it).

\begin{figure}
    \centering
    \includegraphics[width=0.5\linewidth]{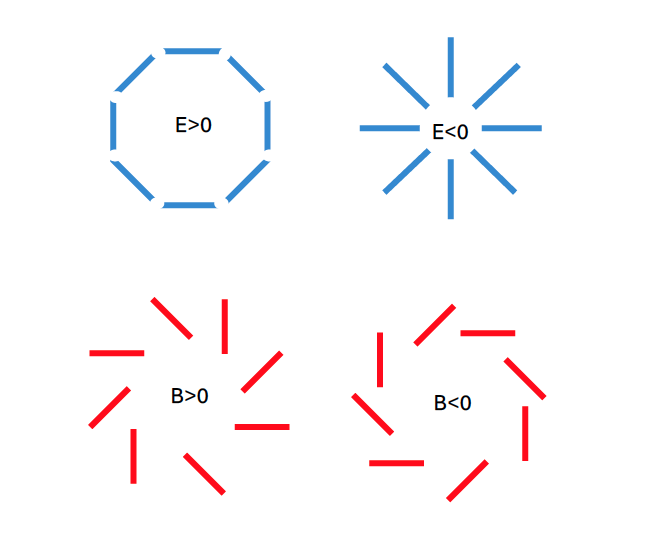}
    \caption{The local polarization pattern near a positive or negative $E$-mode source (top), or a positive or negative $B$-mode source (bottom).
    The $E$-mode pattern is either tangential or radial around the source. As should be expected for even parity, the patterns are unchanged by  taking a mirror image.
    The $B$-mode patterns are at 45$^\circ$ to the radial direction. Again given the odd parity, the $B$-mode pattern is not preserved when taking the mirror image.
    \label{fig:EB}}
\end{figure}

For intuition, it is useful to consider what polarization patterns are represented by $E$ and $B$ modes.
For $E$-modes, the polarization directions are either parallel or perpendicular to the gradient of the polarization magnitude (or wave direction).
For $B$-modes, the polarization directions are $\pm45^\circ$ to the gradient of the polarization magnitude.
If one considers a location in the map with a high amplitude of $E$-modes, one will find a polarization pattern that is either tangential or radial around the point (see Fig.~\ref{fig:EB}).
In contrast,  a location with a high amplitude of $B$-modes will show a corkscrew pattern around the point.
The $E$-mode pattern shows mirror symmetry, while the $B$-mode pattern breaks mirror symmetry.


Beyond invariance to translations and rotations, using the $E$- and $B$-mode basis is convenient from a physics standpoint.
Density perturbations at last scattering only produce $T$ and $E$-modes.
In the absence of vector or tensor perturbations, the CMB radiation field at last scattering will have zero $B$-modes.
This is not true of generic polarization sources: gravitational waves, gravitational lensing, or polarized emission from the Milky Way all generate $B$-modes.
As a result, one can separate the problems: measure density perturbations at last scattering with the CMB temperature and $E$-modes, while using $B$-modes to avoid the large sample variance due to density fluctuations while searching for gravitational waves or while measuring gravitational lensing.


CMB power spectra, $C_{\ell}^{XY}$ for $X, Y (\in T, E, B)$ have been the most important observable constraint from measurements of CMB anisotropy.
We have not yet detected any non-Gaussianity in the primordial CMB perturbations, and the distribution of a Gaussian, statistically isotropic field can be fully described by:
\begin{equation}
    < (a^X_{\ell m})^*a^Y_{\ell'm'} > =\delta_{\ell m} \delta_{\ell' m '} C_{\ell}^{XY}
\end{equation}
The CMB power spectra are often reported in $D_\ell = \frac{\ell(\ell+1)}{2\pi} C_\ell$, as this quantity is nearly constant for a scale-invariant power spectrum.
Note that the CMB spectra are expected to be zero for $TB$ and $EB$; the other four spectra ($TT$, $TE$, $EE$, $BB$) have been observationally measured at high signal-to-noise (see Fig.~\ref{fig:cmbpol_ps} for a compilation of recent measurements).

\subsection{Temperature-Polarization Correlation}

\begin{figure}
    \centering
    \includegraphics[width=0.8\linewidth]{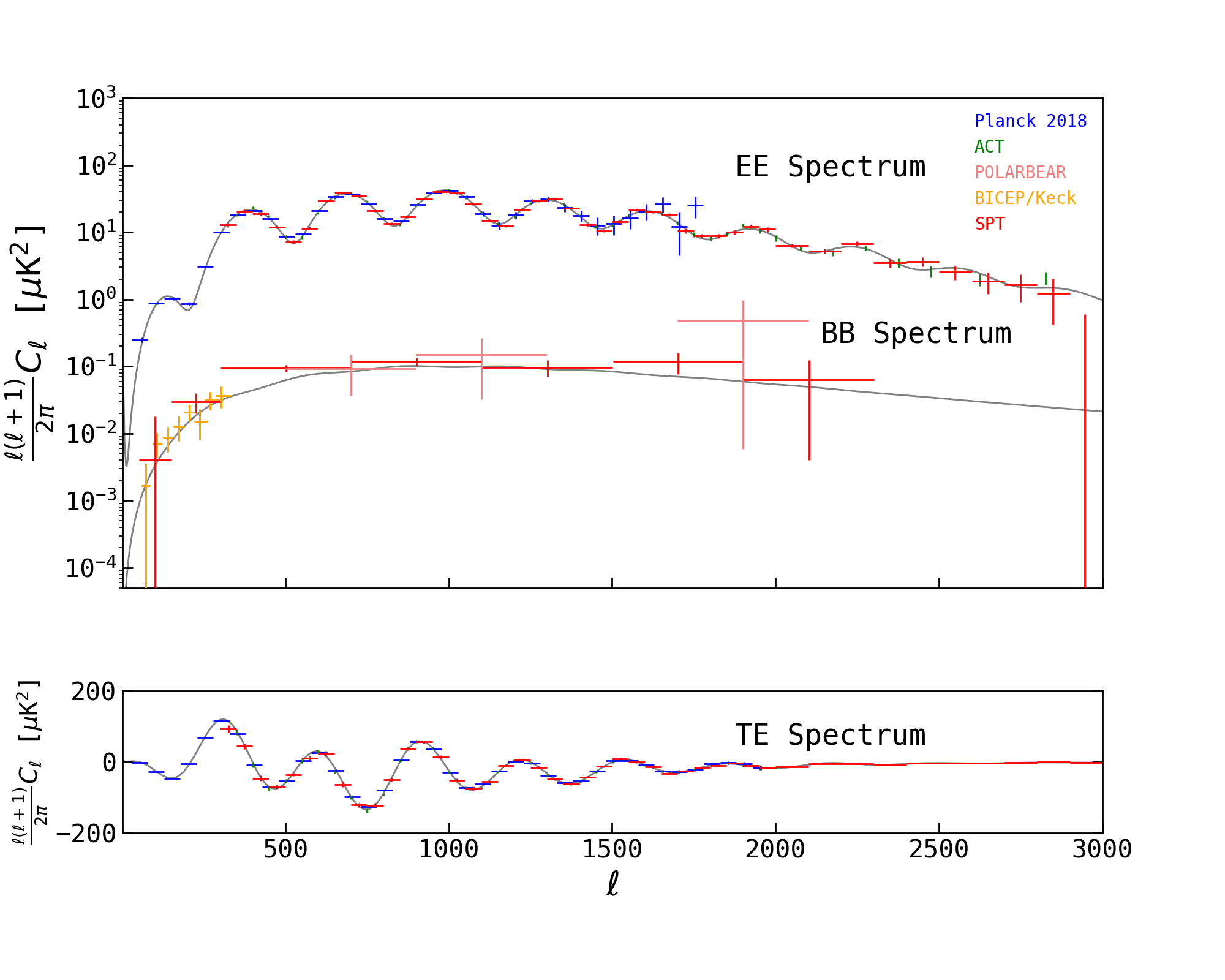}
    \caption{Recent measurements of the CMB polarization power spectra from the Planck satellite \citep{planck2018_5}, ACT \citep{2020JCAP...12..045C}, POLARBEAR \citep{Adachi_2020}, and South Pole Telescope (SPT, \citealt{2020PhRvD.101l2003S, 2021PhRvD.104b2003D}) experiments.
    The $E$-mode auto-power spectrum, $TE$ cross-power spectrum and $B$-mode auto-power spectrum have all been detected at high signal-to-noise.
    }
    \label{fig:cmbpol_ps}
\end{figure}

As acoustic waves in the early hot plasma produce both temperature and $E$-mode polarization anisotropies, it should be unsurprising that there are non-zero correlations between temperature and $E$-mode polarization. 
Measurements of the $TE$ cross power spectrum are useful  for consistency checks and as a source of complementary information to $TT$ power spectrum measurements.
Sources of astrophysical emissions are generally less polarized than the $\sim$10\% polarization of the CMB.
Measurements of polarization, including the $TE$ cross-power spectrum are thus less affected by possible foreground contamination especially at small angular scales \citep{2017A&A...602A..41C}.
This allows the CMB polarization power spectra to be measured to higher multipoles before foregrounds become important.
The $TE$ spectrum has been measured on a wide range of angular scales, from many-degrees to arcminute scales, with selected recent measurements shown in the lower panel of Fig.~\ref{fig:cmbpol_ps}.

\section{Detecting CMB Polarization}\label{chap1:sec5}%

The CMB polarization anisotropies have been measured from the ground, from balloons and from space.
The first detection of CMB polarization was a 4.9\,$\sigma$ detection of  CMB $E$-mode power by the DASI experiment in 2002 \citep{kovac02}.
The  CMB B-modes created by gravitational lensing were first detected in 2013 at 7.7\,$\sigma$ by cross-correlating  $B$-modes measured by SPTpol with a $B$-mode template created using data from the SPIRE satellite and SPTpol \citep{hanson13}.
An internal detection of lensing $B$-modes was first reported at 2.2\,$\sigma$ in 2014 using the POLARBEAR experiment  \citep{ade14}.
Today both the $E$- and $B$-mode power  spectra have been measured at high significance by multiple experiments (see Fig.~\ref{fig:cmbpol_ps}).

\subsection{Observing frequencies and sites}

Observations of CMB anisotropy reach the highest signal-to-noise at observing frequencies near 100-150 GHz (wavelengths 3 - 2 mm).
 This frequency range is also where the flux from the CMB is highest relative to  the flux from other Galactic and extragalactic emission sources.
Synchrotron (radio) emission falls with frequency while thermal dust emission rises with frequency.
The optimal frequency range tends to be slightly lower for all-sky surveys compared  to small-area surveys that are able to target regions with lower Galactic dust emission.
Experiments also include detectors at other frequencies to leverage the different spectral energy distributions to disentangle the CMB anisotropy from the Galactic and extragalactic foreground emission.

\begin{figure}
    \centering
    \includegraphics[scale=0.4]{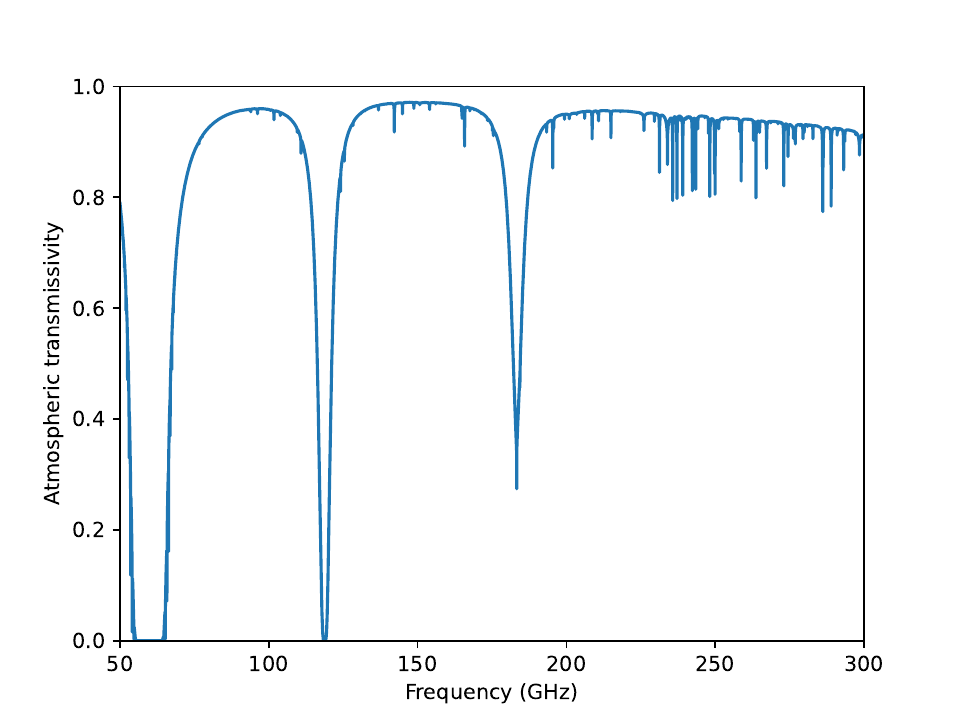}
    \caption{Optical transmission through the atmosphere at the South Pole, estimated using the AM atmospheric model\citep{AM}.
    The major absorption lines are due to water vapor (183 GHz) and molecular oxygen (60 and 118 GHz).
    These three lines explain why ground-based experiments typically define observing bands centered near 95 GHz, 150 GHz and 220 GHz.
    The Atacama plateau in Chile will have less oxygen (being higher elevation) and more water.
    }
    \label{fig:opacity}
\end{figure}

Terrestrial telescopes must also deal with the Rayleigh-Jeans tail of thermal emission from the atmosphere.
The main sources of atmospheric opacity include oxygen (mitigated by high elevation sites) and water vapor (mitigated by very dry and stable sites).
These considerations motivate the two preeminent ground sites: the Atacama plateau at $\sim$5000\,m in Chile and the South Pole at $\sim$3000\,m in Antartica.
Atmospheric emission becomes a very significant source of noise at higher frequencies due the $\nu^2$ frequency scaling of the Rayleigh-Jeans tail and increasing atmospheric opacity (see Fig.~\ref{fig:opacity}).
Space and balloon experiments above all or nearly all of the atmosphere have significant advantages in sensitivity per detector, especially as one moves towards larger angular scales or higher frequencies.
However ground-based experiments can often accumulate significantly more detector-years of data.
Most atmospheric emission is unpolarized, but polarization has been detected from the emission and scattering of atmospheric ice crystals \citep{takakura19,coerver24}.

\subsection{Common hardware elements to detect CMB polarization}

While the design details vary, CMB polarization experiments all need to solve similar problems to measure the ultrafaint CMB polarization anisotropies against a background of instrumental noise, thermal emission from nearby warm objects (like the telescope mirrors), and astrophysical foregrounds.
Each experiment generally includes the following elements:

\begin{itemize}
\item Detectors. Low noise detectors are critical to measuring the very faint CMB polarization anisotropy.
The dominant detector technology is a bolometer (see \S\ref{subsec:bolometer}), which uses a transition-edge sensor to measure the incoming optical power.
At higher frequencies microwave kinetic inductance detectors (MKIDs) are sometimes used for their simpler fabrication.
As one moves towards radio frequencies, interferometry becomes competitive for sensitivity.

\item Cryogenics. Bolometers and MKIDs are operated at sub-Kelvin temperatures, as lower temperatures increase their sensitivity.
Some detector noise terms (e.g., phonons, Johnson noise) increase with temperature.
Active cooling is critical to achieve these temperatures.
Additionally, cooling filters, lenses and other optical elements  reduces the photon noise contribution from their thermal self-emission.
 Ground-based experiments use pulse-tube coolers (PTCs) to cool higher temperature stages (down to $\sim$4\,K).
The PTCs work best when oriented vertically, which can impose operational constraints on observing elevations.
For the detector stage, the field is transitioning from Helium sorption refrigerators which can achieve a few hundred mK with a duty cycle of order 70\%, to more expensive dilution refrigerators which can provide continuous cooling to under 100 mK.
Thus dilution refrigerators can increase observing time while reducing the detector noise levels.
The Planck satellite demonstrated the first use of a dilution refrigerator in space.

\item Polarization-sensitive antennae. This might be a dipole antenna (e.g., \citealt{austermann12,Ade_2017}), a phased array of dipole antennae (e.g., \citealt{Ade_2015}), a sinuous wave antenna (e.g., \citealt{suzuki16, Sobrin_2022}), among other possibilities.
This allows the experiment to detect polarization.

\item (Optional) Polarization modulation. Modulating the polarization signal through means such as a rotating half-wave plate (see highlighted box) can help with noise and the control of instrumental systematics. However, moving parts in the optical path have their own risks.

\begin{BoxTypeA}[chap1:box1]{Half-wave plates}

Certain crystalline materials, such as sapphire, are birefringent, meaning the material has a different index of refraction for  linear polarizations of light depending on the orientation of the linear polarization to the crystalline structure.
The birefringent properties of these crystals can be used to construct half-wave plates (HWPs) such that a phase shift of $\pi$ is induced between the two polarization axes.
This has the effect that if a linearly polarized plane wave enters the HWP polarized at an angle $\theta$ relative to the fast axis of the HWP, the linear polarization of the exiting wave will be at angle $-\theta$, a rotation of $2\theta$.
HWPs in CMB experiments may be cooled to reduce their own thermal emission and include anti-reflection coatings to improve optical transmission.
HWPs are used in CMB experiments to modulate the CMB polarization signal.

\subsection*{Signal modulation}

A HWP rotating at frequency $\nu$ will rotate a static linear polarized source at $4\nu$, i.e. if the HWP is rotating at 1 Hz, a linearly polarized DC signal will now be modulated at 4 Hz.
This can be useful for two reasons.
First, it means both linear polarizations can be measured with the same detector, which can be useful for controlling instrumental systematic errors that can arise from differences between the properties of individual detectors.
Second, it shifts the frequency range  of the polarization signal to higher frequencies which reduces the penalty from low-frequency noise sources, such as atmospheric variability.
\end{BoxTypeA}

\item Lenses and mirrors. These optical elements couple the antennae to the sky and set the size of the beam on the sky.
Experiments targeting the inflationary gravitational wave signal, which can be done with degree-scale beams, frequently use cryogenically cooled lenses only.
Other science goals benefit from few-arcminute-scale beams and include larger diameter mirrors in the optical design.
Baffles and ground shields are frequently added to intercept stray light that might otherwise reach the detectors from the thermal emission of the warm ground .

\item Filters. Filters are designed to allow some frequencies of light to pass while rejecting  frequencies away from the desired signal band.
Filters may achieve this rejection by absorption or reflection.
 Filters are needed for two reasons.
First, blocking infrared light reduces the thermal load on the cryogenic system.
This allows the detector and optical elements to be cooler (reducing their self-emission and thus contribution to the detector photon noise).
Second, while the antenna are sensitive to a broad spectrum of photons, it is useful to define narrower observing frequency bands in order to leverage the different spectral energy distribution of the CMB versus foreground signals like thermal dust emission from the Milky Way to separate these signals.

\end{itemize}

\begin{figure}[htbp]
\centering
\subfloat[Thermal model for bolometer]
{\includegraphics[width=0.5\textwidth]{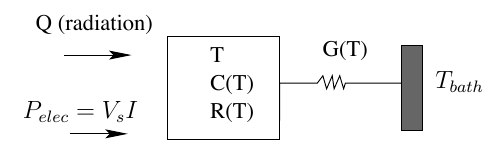}
\label{fig:bolomodel}}
\subfloat[Illustrative R(T) curve for superconductor transition]
{\includegraphics[width=0.5\textwidth]{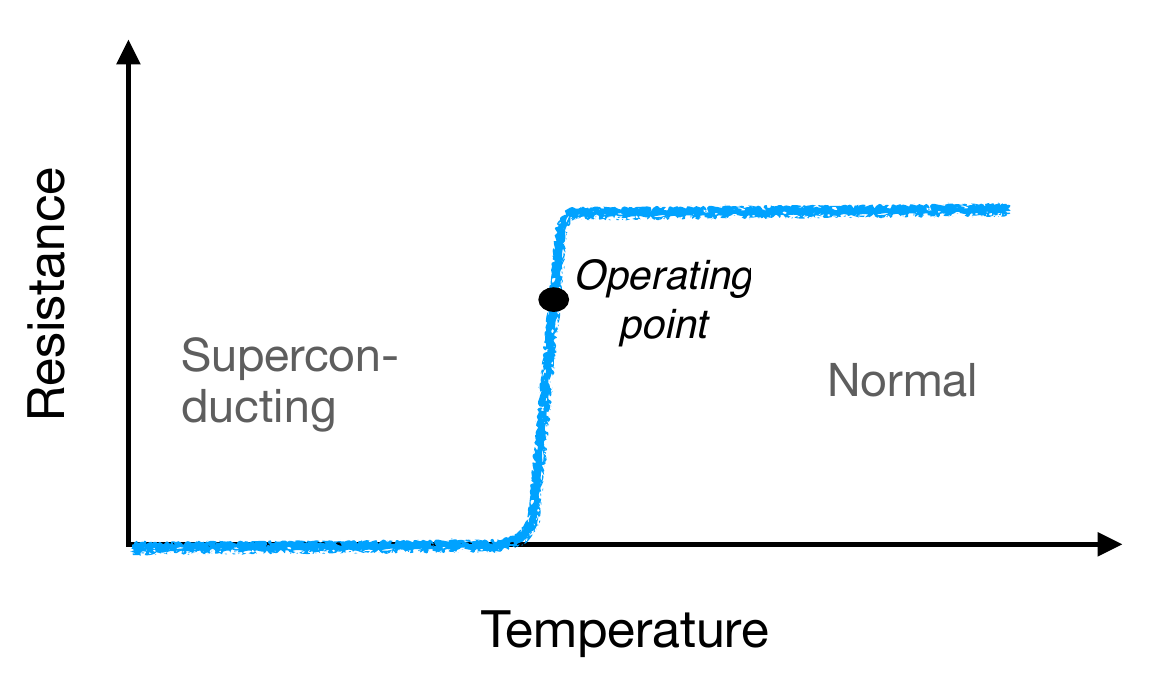}
\label{fig:transition}}
\caption{\textit{Left panel:} A simplified thermal model for a bolometer. The bolometer is heated to temperature T by the incoming power, a combination of radiative power Q and electrical power $P_{elec}$.
At temperature T, the bolometer has a heat capacity $C(T)$ and resistivity $R(T)$.
 It is weakly thermally coupled with thermal conductance $G(T)$ to the bath at temperature $T_{bath}$.
 With electrical power $P_{elec}$ fixed, the bolometer's temperature will increase when radiative power increases (and vice versa), in turn changing the resistance $R(T)$.
 \textit{Right panel:} An illustrative diagram for the resistance as a function of temperature of a superconductor through the transition from superconducting to normal states.
 TES bolometers are operated at a temperature in the transition, marked by the black dot, where there is a very steep change in resistance with temperature.
A SQUID amplifier is used to measure changes in the current flowing through the TES bolometer, which can be calibrated back to changes in incoming optical power.
}
\label{fig:bolometer}
\end{figure}

\subsection{Bolometers}\label{subsec:bolometer}




Bolometers are designed to measure radiative power using a device with a temperature-dependent resistance.
Bolometers are most sensitive when the resistance changes very quickly with temperature: the magnitude of $dR/dT$ is large.
The basic concept is illustrated in  Fig.~\ref{fig:bolomodel}.
The bolometer absorbs radiation (Q) and electrical power ($P_{elec}$).
In steady state, the incoming power is balanced by heat loss through the thermal link $G(T)$ to a colder base at $T_{bath}$.
If the incoming power ($Q+P_{elec}$) increases, the bolometer temperature will increase and change its resistance $R(T)$.
As a CMB telescope scans the sky, the small anisotropies in the CMB will translate to small variations in the bolometer temperature and resistance as a function of the telescope position and the detector's polarization angle.

While bolometers are the mainstay of CMB experiments today,
many earlier experiments, including the DASI experiment that first detected CMB polarization and the WMAP satellite, used interferometric observations \citep{readhead04}.
Inteferometry is are still used at lower frequencies, including the low frequency instrument (LFI) on the Planck satellite.
At higher frequencies, microwave kinetic inductance detectors (MKIDs) are beginning to compete with bolometers \citep{mkid1,mkid2}.
These other detector technologies become especially relevant for expanding the frequency range of observations to better discriminate between the CMB and astrophysical foregrounds (e.g. galactic synchrotron emission is brighter at low frequencies, while galactic dust dominates at high frequencies).

\subsubsection{Neutron transmutation doped Germanium (NTD-Ge) bolometers}

The first generation of bolometers that achieved photon-noise limited performance used semiconducting Germanium thermistors modified with neutron transmutation doping (NTD-Ge bolometers).
These thermistors were suspended on silicon spiderwebs, strategically coated with gold to absorb light, placed behind feedhorns.
The spiderweb structure allowed good absorption, possible polarization sensitivity,  low thermal mass and a low cross-section to cosmic rays.
CMB polarization experiments using these detectors include the first instrument of BICEP (ground), MAXIPOL (balloon), BOOMERanG (balloon), and the Planck satellite (space). 

While individually sensitive, NTD-Ge bolometers are difficult  to scale to larger detector arrays.
The TES bolometers discussed next have three important advantages over the NTD-Ge bolometers.
First, the NTD-Ge bolometers could not be lithographed as part of a monolithic array, but required assembly.
Individually mounting thousands of NTD-Ge devices would be difficult and prone to poor yield.
Second, NTD-Ge bolometers had very high impedance, as opposed to the O(1) \si{\ohm} of TES bolometers.
High impedance leads to a high sensitivity to microphonic noise, effectively requiring the use of liquid nitrogen and liquid helium instead of pulse tube cryocoolers.
Serious microphonic issues could also be created by vibrations from rotating half-wave plates or rapidly scanning telescopes.
High impedance also ruled out the SQUID multiplexing techniques that can be used with TES bolometers.
Individual wires for kilopixel arrays would conduct an unacceptably high thermal load onto the cold stages.
These difficulties motivated the shift to TES bolometers.

\subsubsection{Transition-edge sensor (TES) bolometers}

Transition-edge sensor (TES) bolometers enabled experiments to grow from $\sim$100 to 1000 detectors in the focal plane (see Fig.~\ref{fig:progress}).
TES bolometers are operated at a selected point (e.g. the black dot in the Fig.~\ref{fig:transition}) in a superconductor's transition from the superconducting state to normal state.
The transition temperature of the device is chosen based on the expected optical power and bath temperature.
As this transition happens over a very narrow range of temperatures, there can be a very large $dR/dT$ in the transition even though the normal resistance is low enough to avoid issues with microphonics.
TES bolometers are voltage biased to operate at a specific resistance, e.g. $0.7 R_N$ for 70\% of the normal resistance $R_N$.
When the optical power increases, the electrical power decreases to hold stable the TES resistance (and temperature).
The optical power falling on the bolometer is inferred from the required electrical power, which is read out by means of a SQUID amplifier.
SQUIDs are very sensitive magnetometers that can be used to measure current.
Multiplexing (reading out multiple detectors on a single wire to the cold stages) can be achieved by either biasing individual detectors at different frequencies, by reading out detectors at different times, or a combination of both temporal and frequency multiplexing \citep{irwin10}.
Some examples of CMB polarization experiments using multiplexed TES bolometers include SPT-3G (ground), BICEP3 (ground), Simons Observatory (ground), and SPIDER (balloon) \citep{sobrin22,bicep,mccarrick21,spider}.
Designs for future experiments such as CMB-S4 (ground, 2030) and LiteBIRD (space, 2032) also include TES bolometers \citep{litebird, cmbs4}.

\subsection{Future technology directions}

\begin{figure}[htbp]
\centering
{\includegraphics[width=0.7\textwidth]{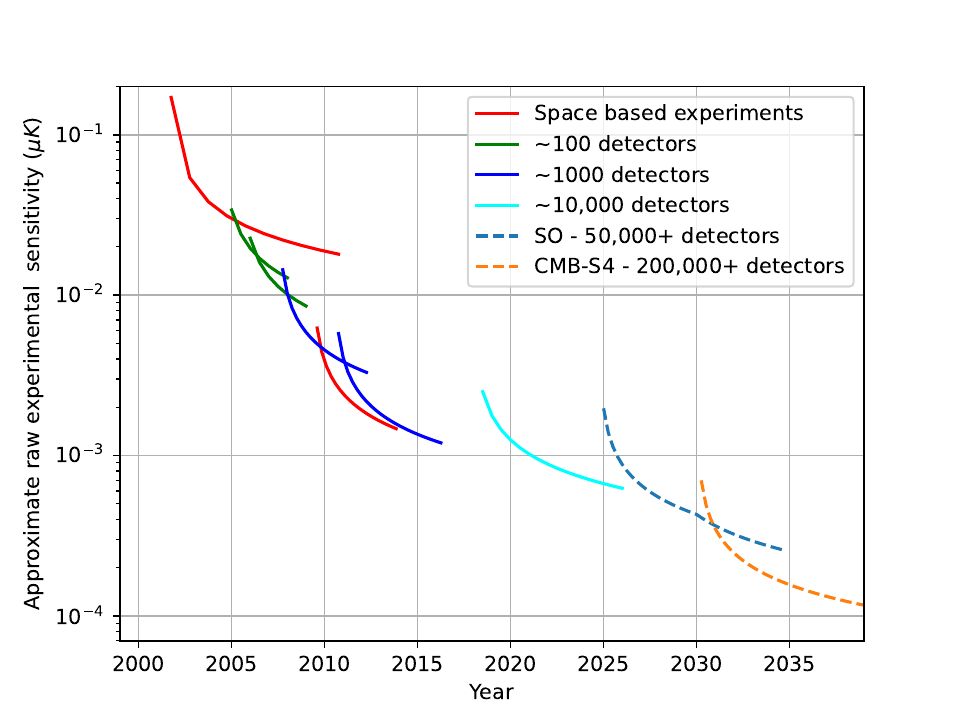}  }
\caption{\label{fig:progress} Progress in raw experimental sensitivity over time. Driven by the ability to construct and field larger detector arrays, the field has seen sensitivity improving at  near-exponential rates with time.
The red lines are for the WMAP and Planck satellites.
The green, blue, and cyan lines are for selected ground-based experiments with O(100), O(1000), and O(10,000) detectors respectively.
The dashed blue line is for the Simons Observatory in Chile, which is taking engineering data in 2024, and has a funded upgrade to approximately double the detector count midway through the 10-year survey.
The dashed orange line is for the planned CMB-S4 experiment, which will observe for an estimated 10 years from Chile and possibly the South Pole.
}
\end{figure}

As the sensitivity of  individual CMB detectors has been limited for some time by the fundamental noise floor set by the randomness of photons arriving at the detector,
much effort has been devoted to building receivers with ever increasing numbers of detectors.
This has led to a version of ``Moore's law" showing near exponential growth in CMB detector counts with time (see Fig.~\ref{fig:progress}).
The continued drive for more detectors has necessitated progress on a number of related technological fronts, such as:
\begin{enumerate}
\item Improving the yield percentage and repeatability on the nanofabrication process of detectors.
\item Redesigning the telescope optics to increase the focal plane area (allowing more detectors on the same telescope).
\item Increasing the maximum manufacturable size for optical filters, lenses, and their anti-reflection coatings (again allowing larger focal plans).
\item Improvements to the detector readout electronics, including work to reduce readout noise, to improve the ease and reliability of biasing detectors, and to increase the multiplexing factor  and thus reduce the thermal load on the cryogenic system.
\end{enumerate}
Very active technological development programs have been crucial to enable each generation of CMB experiment to be noticeably more sensitive.

\section{Conclusions}

Observations of CMB polarization anisotropies are increasingly important to our understanding of the universe.
Upcoming measurements of CMB polarization anisotropies present opportunities to detect inflationary gravitational waves, to discover new light particles produced during the Big Bang, and test the standard model of cosmology.

\begin{ack}[Acknowledgments]

We thank Jessica Zebrowski, Justin Clancy, Nicholas Huang, Srinivasan Raghunathan, Junhao Zhan, and Michael Doohan for the valuable feedback they provided on the draft.

\end{ack}


\bibliographystyle{Harvard}
\bibliography{reference}

\end{document}